\documentstyle[preprint,aps]{revtex}
\begin{document}

%
%

\title{Photoemission spectra of ${\rm Sr_2 Cu O_2 Cl_2}$: a theoretical 
analysis}

\author{A. Nazarenko,$^1$ K.J.E. Vos,$^2$ S. Haas,$^1$ E. Dagotto,$^1$ 
and R.J. Gooding$^2$}

\address{$1.$ Department of Physics and National High Magnetic Field Lab,
Florida State University, Tallahassee, FL 32306, USA}

\address{$2.$ Department of Physics, Queen's University, Kingston,
Ontario, Canada K7L 3N6}

\date{\today}

\maketitle

\begin{abstract}

Recent angle resolved photoemission (ARPES) results for the insulating 
cuprate 
${\rm Sr_2 Cu O_2 Cl_2}$ have provided the first experimental
data which can be directly compared to the (theoretically) well--studied
problem of a single hole propagating in an antiferromagnet. 
The ARPES results reported a small bandwidth,
providing evidence for the existence of strong correlations in the cuprates.
However, in the same experiment some 
discrepancies with the familiar 2D ${\rm t-J}$ model were also observed. 
Here we discuss a comparison between the ARPES results and
the quasiparticle dispersion of
both (i) the ${\rm t-t'-J}$ Hamiltonian 
and (ii) the three--band Hubbard model 
in the strong--coupling limit.  Both model Hamiltonians
show that the experimentally observed one--hole band structure
can be approximately reproduced using reasonable values for ${\rm t'}$, or 
the direct oxygen hopping amplitude ${\rm t_{pp}}$.

\vskip 0.5 truecm

To appear in Phys. Rev. B, April 1st, as a Rapid Communication.
\end{abstract}

\pacs{74.20.-z, 74.20.Mn, 74.25.Dw}

The dispersion of the hole quasiparticles in the normal state of the
two--dimensional (2D) ${\rm Cu O_2}$ planes is of crucial importance in
the search for a microscopic theory of the high temperature superconductors. 
To be specific, comparison between experimentally observed 
and theoretically predicted quasiparticle dispersions allows for the
scrutiny of Hamiltonians proposed to describe the carriers in
these planes.  Recently, angle resolved photoemission (ARPES) 
data \cite {flat-exper} 
have shown that for Bi2212 and YBCO, both near optimal doping, 
the quasiparticle bandwidth is only ${\rm \sim 0.3eV}$. In addition, 
the dispersion contains an interesting flat region around 
${\rm X = (\pi,0)}$ and ${\rm Y=(0,\pi)}$ (in the notation
of the 2D square lattice).  The apparent universality of these results 
among different hole-doped compounds implies that their origin must lie 
in the ${\rm Cu O_2}$ planes. 
The experimentally observed small bandwidth suggests that 
strong correlations are crucial for a proper description of
carriers in the cuprates. Further, calculations based upon the ${\rm t-J}$ 
model \cite {flat} have shown that the flat regions near ${\rm X}$ and 
${\rm Y}$ are naturally obtained for holes moving in 
an $antiferromagnetic$ (AFM)
background, the latter being a direct consequence of the effect of strong 
correlations at half filling.  Then, for doping levels such that the AFM
correlation length, $\xi_{AF}$, is robust,
the dispersion obtained from the 2D ${\rm t-J}$ model should
approximate the observed quasiparticle dispersion.

In a recent letter, Wells et al. \cite {wells} reported
ARPES measurements on ${\rm Sr_2 Cu O_2 Cl_2}$. This material is
an $insulating$ layered copper oxide that is difficult to
dope. These results are extremely important because this experiment 
precisely corresponds to the well--studied theoretical problem of a 
single hole moving in a quantum antiferromagnet, and thus is a
problem that serves as a test case for any proposed microscopic Hamiltonian,
as we now discuss. From 
Ref. \cite {wells} the main message for theorists can be summarized in 
two items: (i) on the one hand, the ARPES spectra shows that
the 2D ${\rm t-J}$ model accurately describes the results along the
diagonal $(0,0)$ to $(\pi,\pi)$ in momentum space. The state with the 
highest 
energy in the valence band is obtained at ${\rm {\bar M} = (\pi/2,\pi/2)}$ 
as 
expected from many calculations \cite {review}.
Further, the bandwidth of the quasiparticle is very close to the 
bandwidth obtained for one hole in the 2D ${\rm t-J}$ model using various 
techniques \cite {review,liu}. This result is
of considerable importance for theories of high--${\rm T_c}$ as it supports
the use of one--band models to describe the electronic properties of
the cuprates; (ii) on the other hand, Wells et al. \cite {wells} noticed 
that the 2D ${\rm t-J}$ model prediction fails near ${\rm X = (0,\pi)}$ 
since the experiments
do not confirm that the energies of the ${\rm {\bar M}}$ and ${\rm X}$ points
are nearly degenerate, as is predicted by ${\rm t-J}$ model 
calculations \cite {review,liu}.

It is the purpose of this paper to provide evidence that item (ii) above
can be naturally accounted for in two different ways, and we shall discuss
each of these ideas separately. Firstly, if the ${\rm t-J}$ model is 
enlarged 
to contain hole hopping amplitudes at distances larger than one lattice
spacing, one can indeed fit the ARPES data.
There is no symmetry constraining the 
hole hopping to just nearest-neighbors, and in several studies it has been 
shown
that to reproduce the spectra of three--band Hubbard models using one--band 
${\rm t-J}$
models on small clusters, it is necessary to include  
a ``${\rm t'}$-hopping'' along 
the plaquette diagonals \cite {hyb}. As long as the ratio
${\rm |t'/t|}$ is small, this term does not affect the AFM
properties of the model near half-filling, since this hopping only connects
sites that are on the same sublattice of a N\'eel ordered state. However,
such terms will clearly affect the quasiparticle dispersion.

It has, in fact, become standard to use Hamiltonians including ${\rm t'}$ 
hoppings 
to describe several compounds. Some authors \cite {maekawa} have argued that
to fit the ARPES Fermi surface of different materials it is necessary to
select ${\rm t'=-0.2t}$ for ${\rm La_{2-x} Sr_x Cu O_4}$; ${\rm t'=-0.45t}$ 
for ${\rm YBCO}$; and ${\rm t'=+0.2t}$ for ${\rm Nd_{2-x} Ce_x Cu O_4}$,
and there seems to be some justification for this approach. 
For example, it was recently shown \cite {ehasym} that the differing
magnetic properties of the LaSrCuO and NdCeCuO systems could be accounted 
for 
{\em only} if such a next--nearest--neighbour hopping term was included.
Since all these compounds have the same ${\rm Cu O_2}$ planes, the 
strength of 
these effective ``${\rm t'}$-terms'' changes among the different cuprates
likely due to three dimensional effects (bi-layer hopping, influence of 
apical oxygens,
influence of orbitals usually neglected in the 3 and 1 band descriptions 
of the
cuprates, etc.).  Then, it is reasonable to expect that the ${\rm t-t'-J}$ 
model can mimic the dispersion of ${\rm Sr_2 Cu O_2 Cl_2}$ with a properly 
selected amplitude ${\rm t'}$. Although it is clear that fitting data with 
an arbitrary parameter is
less satisfying than microscopically deducing the value of ${\rm t'}$,
in the absence of a microscopic theory to compute ${\rm t'}$
the best we can do is show that a reasonable value of this
amplitude leads to agreement with the experiments. Note that we only have 
one 
free parameter to fit an entire function of momentum (dispersion), and thus 
not much extra freedom is introduced in the model with a ${\rm t'}$ 
amplitude.

The ${\rm t-t'-J}$  model Hamiltonian is defined as
$$
{\rm H  =  
~-~t \sum_{ \langle {\bf i}{\bf j} \rangle }
({\bar c}^{\dagger}_{{\bf i}\sigma} 
{\bar c}_{{\bf j}\sigma} + h.c. )
-~t'~\sum_{ \langle\langle {\bf i}{\bf i{^\prime}} \rangle\rangle }
({\bar c}^{\dagger}_{{\bf i}\sigma} 
{\bar c}_{{\bf i^{\prime}}\sigma} + h.c. )} $$
$$
{\rm~+~J \sum_{ \langle {\bf i}{\bf j} \rangle }
( {\bf S}_{{\bf i}} \cdot {\bf S}_{{\bf j}} - 
{{1}\over{4}} n_{\bf i} n_{\bf j} ),  }
\eqno(1)
$$
\noindent 
where ${\bar c}$ are hole operators, $\langle \rangle$ refers to n.n.
sites, $\langle\langle\rangle\rangle$ refers to n.n.n. sites (along the 
diagonal of the plaquettes), and the rest of the notation is standard. 
Using the Born approximation (extensively discussed in the literature 
\cite {rainbow,liu})
the dispersion of one hole in the ${\rm t-t'-J}$ model was here calculated
for several values of ${\rm t'}$ \cite {comment}.
To select ${\rm t'}$ the experimentally observed $isotropy$ of the 
dispersion of ${\rm Sr_2 Cu O_2 Cl_2}$ near the top of
the band at ${(\pi/2,\pi/2)}$ \cite {wells} was 
used. We followed this criterion
since it is the most clearly defined feature in the data of 
Ref. \cite {wells}
(concentrating our effort on the measured behavior near ${\rm X}$ would be 
dangerous since for these momenta the ARPES data only provide a bound on the
quasiparticle energy). We observed that to reproduce the 
${\rm (\pi/2,\pi/2)}$ isotropy, an amplitude ${\rm t' \sim - 0.35 t}$ is
necessary (supplemented by ${\rm J=0.125 eV}$ as energy scale and a ratio
${\rm J/t}$ around 0.3--0.4, as is usually assumed). This value of ${\rm t'}$
is similar to that found in earlier comparisons based on small cluster 
exact 
diagonalization results \cite {ehasym} to the ARPES data \cite {ole}.
The resulting dispersion is shown in Fig. 1, and is compared
with the experimental data. Also shown is the dispersion produced by the 
${\rm t-J}$
model. The agreement near the top of the band is much improved by the inclusion
of ${\rm t'}$, especially in the ${\rm (0,\pi) \to (\pi,0)}$ direction. 
It is clear that along this direction the ${\rm t-J}$ and 
${\rm t-t'(=-0.35t)-J}$ 
models have substantially different dispersion relations.
On the other hand, along the ${\rm (\pi,0) \to (0,0)}$ line our results have 
more structure than the reported experiments. However, experimentally along 
this line it 
is not possible to extract a reliable quasiparticle dispersion 
(see Fig. 2a of Ref. \cite {wells}) and thus we do not consider this 
disagreement with states having an energy much less
than the chemical potential a serious problem. Actually,
it is possible that only the quasiparticles near ${\rm \bar M}$ are
relevant for normal state and superconducting properties \cite {tc}.  
Nevertheless, it is important to fix this quantitative discrepancy, and for 
this purpose we turn to our second technique for producing a quasiparticle
dispersion that agrees with the data of Ref. \cite {wells}.

The underlying physics of carrier transport in weakly doped ${\rm Cu O_2}$ 
planes 
is governed by oxygen hole motion in an AFM background, and thus one might
expect that the three--band Hubbard model in the strong--coupling limit 
\cite {emery} 
should be employed. However, as is now well known, a renormalization
procedure approximately  maps the low--energy physics of the three--band 
Hubbard model 
into that of the one--band ${\rm t-J}$ model \cite {zhangrice}. Subsequently,
a perturbative calculation \cite {shastry} prescribed conditions under
which further--than--nearest--neighbour hopping terms must be included. As 
demonstrated 
above, if one wishes to fit the ARPES data of Ref. \cite {wells} 
{\em at least} the next--nearest--neighbour hopping term ${\rm t'}$ is 
required. 

One possible caveat to the reduction to one--band models has been discussed 
in previous 
papers \cite {zhangrice,shastry,frenkel}. To be specific, the reduction 
from the three--band Hubbard model to the ${\rm t-J}$ model
eliminates the spin degree of freedom of the oxygen carriers,
effectively producing spinless vacancies. However, if the direct
oxygen--oxygen hopping is large enough, the oxygen hole can acquire a 
non--zero spin. 
As a demonstration of the potential inadequacies of any renormalization to
a one--band model based on carriers possessing no spin, consider firstly 
that {\em unlike}
the ${\rm t-t'-J}$ model without the ${\rm t'}$ term, holes described by 
the three--band model moving in a {\em rigid} Ising-like AFM are mobile 
\cite {roth}, and
thus these carriers possess a spin degree of freedom. However, in order to 
enhance the delocalization 
of these holes in a quantum AFM, the spins become distorted, and the spin 
of the carrier
is quenched \cite {zhangrice,frenkel}, approximating the spinless vacancies 
of the one--band
${\rm t-t'-J}$ model. If one then includes a hopping process that is 
{\em independent}
of the magnetic background, such as a direct oxygen-oxygen hopping term, 
and the energy scale
of this additional hopping term is larger than any of the other terms, the
spin of the carrier is no longer quenched \cite {frenkel}, and one should 
then
not expect that such carrier motion (in a strongly correlated magnetic 
background) should be describable by the spinless vacancies of the 
${\rm t-t'-J}$ model.

Motivated by this observation, we have examined the usefulness of fitting 
the 
experimental results with the microscopic Hamiltonian of 
Ref. \cite {frenkel}.
We have studied the analogue of the three--band ${\rm t-J}$ model found 
from the
strong coupling limit of the three--band Hubbard model. This is a complicated
microscopic Hamiltonian, and we refer the reader to Ref. \cite {frenkel} for
a detailed discussion of Eq. (2) of that paper. In addition to this, here
we append the Hamiltonian of Ref. \cite {frenkel}
with a direct oxygen-oxygen hopping term, something that
we found to be crucial to our success in fitting the ARPES 
data \cite {wells}.
The parameter values for this Hamiltonian \cite {3bandcomment} can be
estimated from the parameters of the three--band Hubbard 
model \cite {hyb}, with the ${\rm Cu-O}$ exchange interaction reduced by 
a direct exchange term \cite {george}.  Then, following the notation of 
Ref. \cite {frenkel}, one finds
$$
{\rm J_1 = 0.032,~J_2 \approx 0.16,~t_a = 0.38,~t_b = 0.43,}$$
$${\rm ~and~t_{pp}=-0.65}
\eqno(2)
$$
Extrapolating the analysis of Shastry \cite {shastry} to include the 
direct oxygen 
hopping, and according to the above argument concerning the potential 
inadequacies of the 
renormalization to a one--band ${\rm t-t'-J}$ model, the important ratio of 
hopping 
parameters is ${\rm {t_{pp} / | t_a+t_b |}~\approx~0.8}$.
This implies that the renormalization to spinless vacancies may not be 
entirely justified.

We have used these parameters in an exact diagonalization study of the
Hamiltonian of Ref. \cite {frenkel} for a $4 \times 4$ ${\rm Cu O_2}$ 
cluster with periodic boundary conditions.  Our results are shown in Fig. 2.
Note that the shape of the quasiparticle dispersion (at least for the small
number of wave vectors that we can access) is identical to that of Wells'
data, which is encouraging. However, the bandwidth is roughly half of that 
found experimentally. This latter result is not that surprising given that 
for a large ${\rm t_{pp}}$, important finite--size effects are to be 
expected \cite {gagliano}.

In order to better assess the quasiparticle dispersion relation that 
describes a single oxygen
hole doped into an AFM that follows from Eq. (2) of Ref. \cite {frenkel}, 
we have
generated the quasiparticle dispersion using a semiclassical variational 
wave function that {\it includes} the spin fluctuations of
the Oxygen holes
where the distortions of the $Cu$ spins away from AFM order are determined 
variationally (see Eq. (19)
of Ref. \cite {frenkel} for a version of this wave function
that does {\it not} include Oxygen spin fluctuations). 
We expect that this semiclassical representation will lead to an excellent 
approximation to the true quasiparticle dispersion quite simply
because the carrier motion has a large direct oxygen hopping. 
(For example, if
one modifies the results of Roth \cite {roth} to include ${\rm t_{pp}}$ and 
allows for the hopping parameters of Eq. (2) to become fitting parameters, 
even if one ignores the Oxygen hole spin fluctuations
one can again fit the Wells' data (shape as well as bandwidth) reasonably 
well --- this is also displayed in Fig. 2.)

We have included these spin distortions \cite {SS,frenkel}, and numerically
solved the associated variational problem to obtain the minimum energy state 
at each wave vector.
This calculation has been carried out for an infinite lattice.
Here, we have again
allowed for the hopping parameters of Eq. (2) (above) to be variables that 
we use to fit 
the quasiparticle dispersion found experimentally, expecting that we obtain 
values close to 
those of the perturbation theory \cite {frenkel}.  Our results, for 
${\rm t_a \sim 0.0,~t_b = 0.3,~and~t_{pp} = -0.35}$, 
are shown
in Fig. 3. Clearly, the agreement with experiment is quite 
good, and improves that
found using the ${\rm t-t'-J}$ model along the $(0,0) \to (0,\pi)$ branch. 
This may point 
to the importance, at least for these crystal momenta, of the spin degree of 
freedom of the holes.

Wells et al. \cite {wells} made the apparently reasonable assumption of 
grouping
together the results obtained for hole--doped ${\rm Bi2212}$ 
\cite {flat-exper},
where flat bands at $(0,\pi)$ near the Fermi energy were reported, 
with those for insulating ${\rm Sr_2 Cu O_2 Cl_2}$, concluding
that as a function of doping states near $(0,\pi)$ must either be
created or move up in energy in a {\em non-rigid} way.  This is one
possible interpretation. However, the many clear differences in the
ARPES Fermi surfaces of different high--T$_c$ compounds leads us to believe
that it is not correct to group Bi2212 with ${\rm Sr_2 Cu O_2 Cl_2}$
since these compounds differ in microscopic details that are relevant 
for quantitative comparisons --- e.g., the numerical value of the 
${\rm t'}$ amplitude. While it is likely that the qualitative mechanism of
superconductivity is the same in all hole-doped cuprates, their 
differences in critical temperature and ARPES data must be caused by 
non--universal features. Thus, in our opinion, 
the data  from Bi2212 and ${\rm Sr_2 Cu O_2 Cl_2}$ cannot be grouped
together to obtain conclusions about the behavior of correlated
electrons {\em as a function of doping} in the cuprates.
To clarify this important point it will be necessary to carry out 
photoemission experiments on the $same$ compound at different densities of 
carriers.

Summarizing, using a ${\rm t-t'-J}$ model it is possible to fit 
the ARPES quasiparticle dispersion data for ${\rm Sr_2 Cu O_2 Cl_2}$ 
reasonably well at the top of the valence band. To obtain an even better 
quantitative agreement the use of a strong--coupling three--band Hubbard 
model 
is required \cite {frenkel}.

We thank B. Wells, R. Laughlin, S. Maekawa, V. Emery and G. Reiter for 
useful 
discussions.  A.N. is supported by the
National High Magnetic Field Laboratory, at FSU.
E.D. is supported by the Office of Naval Research under
grant N00014-93-0495, and the donors of the Petroleum Research Fund
administered by the American Chemical Society. 
S.H. is supported by the Supercomputing Computations Research Institute, 
at FSU. 
K.V. and R.G. acknowledge financial support from NSERC of Canada.

\vfil\eject

%
%
{\bf Figure Captions}
\begin{enumerate}

\item 
Quasiparticle dispersion of the ${\rm t-t'-J}$ model
calculated in the Born approximation for an infinite lattice
using ${\rm t'=-0.35t}$ (solid line), ${\rm J/t=0.3}$ and ${\rm J=0.125
eV}$, compared against the experimental ARPES data (open circles) for 
${\rm Sr_2 Cu O_2 Cl_2}$ of Ref. [3].  The dashed line is the
${\rm t-J}$ model result of Ref. \cite {liu}.

\item 
Quasiparticle energies (stars) for an oxygen hole described by
the effective Hamiltonian of Eq. (2) of Ref. \cite {frenkel} found from
an exact diagonalization study on a $4 \times 4~{\rm CuO_2}$ cluster,
compared against the ${\rm Sr_2 Cu O_2 Cl_2}$ ARPES data (open circles);
the zero of energy of the exact diagonalization numbers is offset 
for clarity.  Also, the solid line denotes the band of energies for hole 
motion in a rigid (viz., Ising like) AFM background 
using hopping parameters ${\rm t_a = 0.12,~t_b = 0.17,~
and~t_{pp} = -0.18}$.

\item 
Quasiparticle energies (solid diamonds) of an oxygen hole in the 
effective Hamiltonian of 
Eq. (2) of Ref. \cite {frenkel} for an infinite lattice using a 
semiclassical variational wave function that includes the
spin fluctuations of the Oxygen carriers.
The quasiparticle dispersion measured in Ref. [3] is also shown 
(open cirlces).
The hopping parameters that produce this quasiparticle dispersion
are ${\rm t_a = 0.0,~t_b =0.3,~and~t_{pp} = -0.35}$.

\end{enumerate}


\begin{references}
\bibitem{comment} We have verified that this approximation produces results 
that are in good agreement with those found from  Green's Function 
Monte Carlo techniques
at the particular value ${\rm t'=-0.2t}$ and ${\rm t'=+0.15t}$ 
(see Ref.[2]). 
Larger ratios of ${\rm |t'/t|}$ lead to severe ``sign-problems'', and 
thus our analysis is carried out with the Born approximation.
\bibitem{3bandcomment} The values of the parameters used in Ref.[14] 
were based
on preliminary estimates for three--band Hubbard model parameters. In what 
follows we use what are believed to be better estimates --- e.g., these 
parameters properly reproduce the Cu--Cu superexchange to be 0.128 eV.
\end{references}
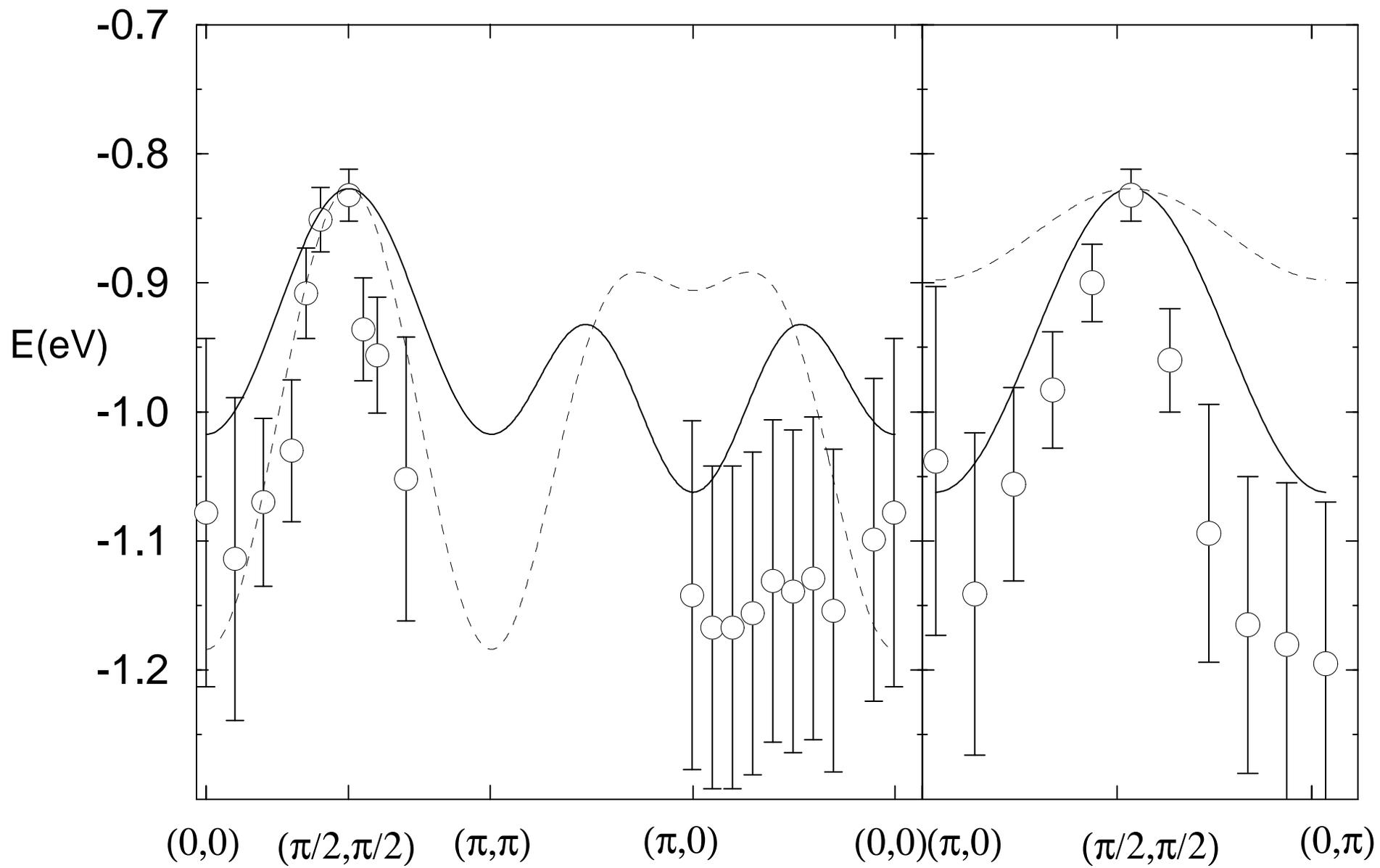
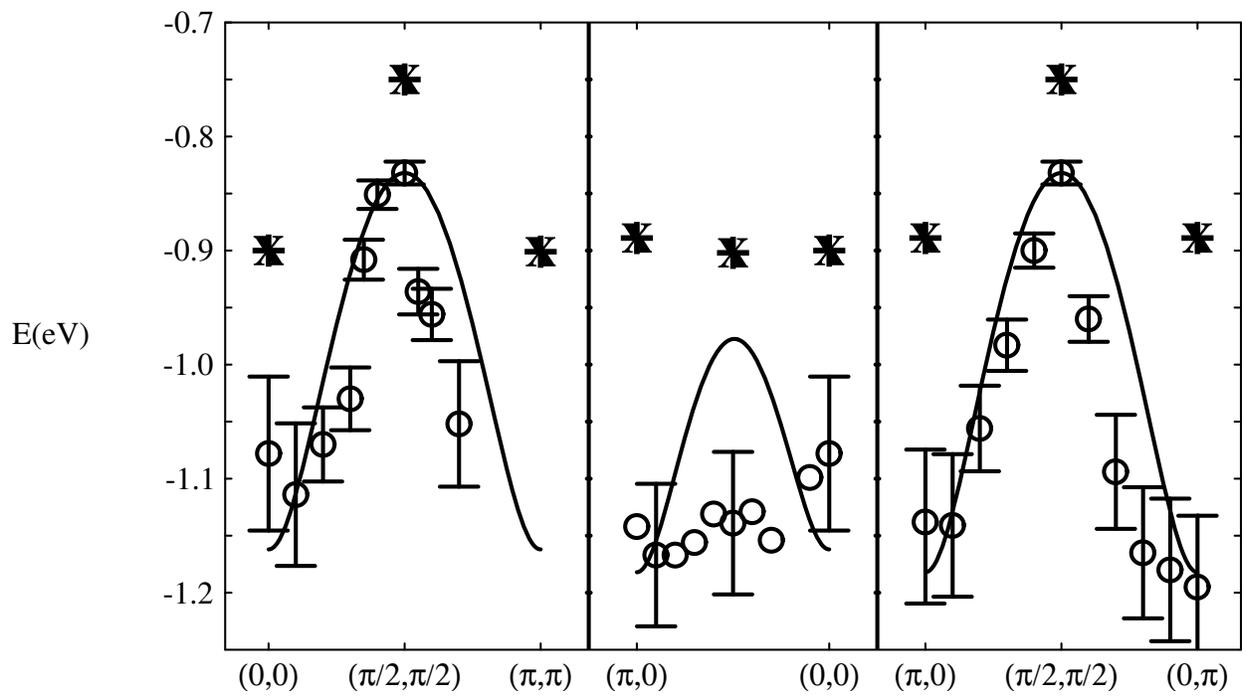
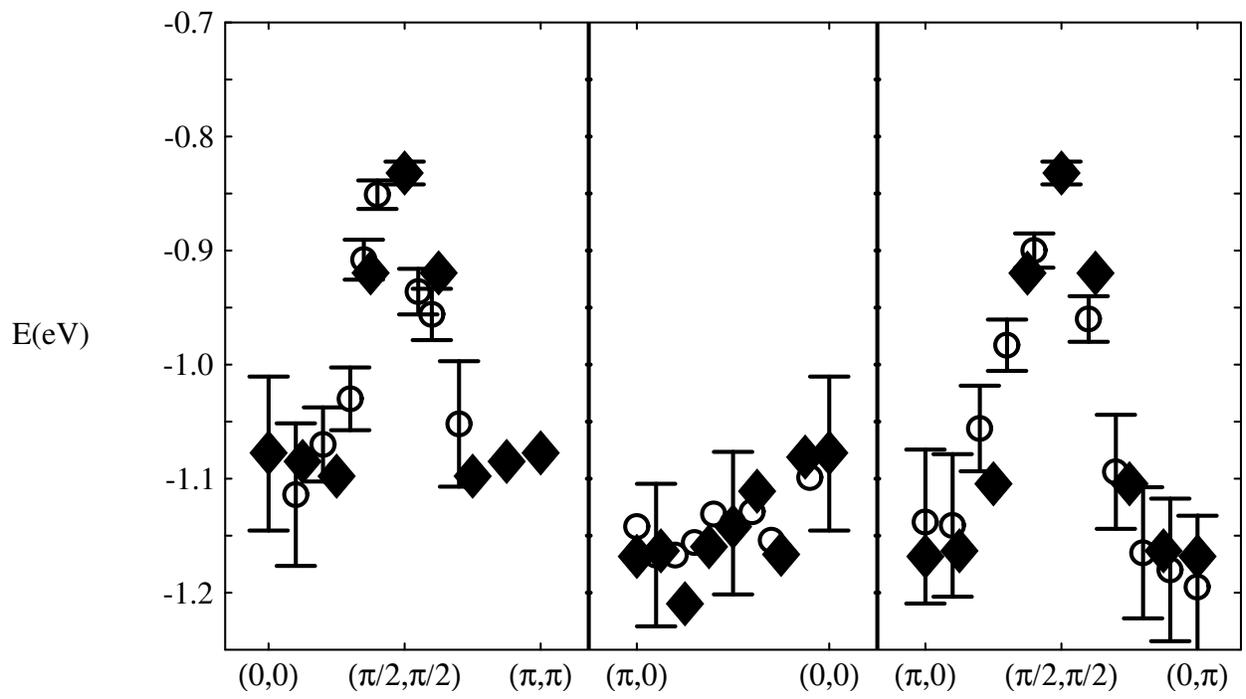
\end{document}